\documentclass[aps,prl,twocolumn,groupedaddress]{revtex4}

\usepackage{amsfonts,amssymb}
\usepackage{amsthm}
\usepackage{tikz}
\usetikzlibrary{calc}
\usetikzlibrary{arrows}
\usepackage{tikz-3dplot}
\usepackage{color} 
\usepackage[fleqn]{amsmath}
\usepackage{amsthm}
\usepackage{amssymb}
\usepackage{amsfonts}
\usepackage{bbm}
\usepackage{epsfig}
\usepackage{times}
\usepackage{hyperref}
\usepackage{bm}
\usepackage{float} 
\usepackage{appendix} 
\usepackage{mathrsfs}
\usepackage{appendix}

\newcommand*{\braopket}[3]{\langle #1| #2 | #3\rangle}

\usepackage{graphicx}

\DeclareGraphicsExtensions{.png}

\newcommand{\comment}[1]{}
\newcommand{\ket}[1]{\left |  #1 \right\rangle}
\newcommand{\bra}[1]{\left \langle #1  \right |}
\newcommand{\braket}[2]{\langle #1|#2\rangle}

\theoremstyle{plain}

\theoremstyle{definition}

\begin{document}

	\title{Summoning, No-Signalling and Relativistic Bit Commitments}

 \author{Adrian \surname{Kent}} \affiliation{Centre for
	Quantum Information and Foundations, DAMTP, Centre for Mathematical
	Sciences, University of Cambridge, Wilberforce Road, Cambridge, CB3
	0WA, U.K.}  \affiliation{Perimeter Institute for Theoretical
	Physics, 31 Caroline Street North, Waterloo, ON N2L 2Y5, Canada.}
	\date{\today}

\begin{abstract} 
Summoning is a task between two parties, Alice and Bob, with
distributed networks of agents in space-time.   Bob gives Alice
a random quantum state, known to him but not her, at some point.
She is required to return the state at some later point, belonging
to a subset defined by communications received from Bob at other
points.   Many results about summoning, including the impossibility
of unrestricted summoning tasks and the necessary conditions 
for specific types of summoning tasks to be possible, follow
directly from the quantum no-cloning theorem and the relativistic
no-superluminal-signalling principle.  The impossibility of
cloning devices can be derived from the impossibility of superluminal
signalling and the projection postulate, together with assumptions
about the devices' location-independent functioning. 
In this qualified sense, known summoning results follow from 
the causal structure of space-time and the properties of quantum
measurements. 
Bounds on the fidelity of approximate cloning can 
be similarly derived. 
Bit commitment protocols and other cryptographic protocols based on 
the no-summoning theorem can thus be proven secure 
against some classes of post-quantum but non-signalling adversaries. 

\end{abstract}
	
		\maketitle
	
\section{Introduction}

To define a summoning task\cite{kent2013no,kent2012quantum}, we consider
two parties, Alice and Bob, who each have networks of collaborating
agents occupying non-overlapping secure sites throughout space-time. 
At some point $P$, Bob's local agent gives Alice's local agent a
state $\ket{\psi}$.   The physical form of $\ket{\psi}$ and the dimension of
its Hilbert space $H$ are pre-agreed; Bob knows a classical description
of $\ket{\psi}$, but from Alice's perspective it is a random state
drawn from the uniform distribution on $H$.    
At further pre-agreed points (which are often taken to all be in the causal
future of $P$, though this is not necessary), Bob's agents send
classical communications in pre-agreed form, satisfying pre-agreed
constraints,  to Alice's local agents, which collectively
determine a set of one or more valid return points.   Alice may manipulate
and propagate the state as she wishes, but must return it to Bob at one of 
the valid return points.   We say a given summoning task is {\it
  possible} if there is some algorithm that allows Alice to ensure
that the state is returned to a valid return point for any valid
set of communications received from Bob.    

The ``no-summoning theorem''  \cite{kent2013no} states that summoning
tasks in Minkowski space are not always possible.
We write $Q \succ P$ if the space-time point $Q$ is in the causal
future of the point $P$, and $Q \nsucc P$ otherwise; we write
$Q \succeq P$ if either $Q \succ P$ or $Q=P$, and $Q \nsucceq P$
otherwise. 
Now, for example, consider a task in which Bob may request at one of two
``call'' points $c_i \succ P$ that the state be returned at a 
corresponding return point $r_i \succ c_i$, where $r_2 \nsucceq c_1$
and $r_1 \nsucceq c_2$.   An algorithm that guarantees that Alice
will return the state at $r_1$ if it is called at $c_1$ must
work independently of whether a call is also made at $c_2$, since
no information can propagate from $c_2$ to $r_1$; similarly if
$1$ and $2$ are exchanged.   If calls were made at both $c_1$ and
$c_2$, such an algorithm would thus generate two copies of $\ket{\psi}$
at the space-like separated points $r_1$ and $r_2$, violating the
no-cloning theorem.   
This distinguishes relativistic quantum theory from both 
relativistic classical mechanics and non-relativistic quantum
mechanics, in which summoning tasks are always possible provided
that any valid return point is in the (causal) future of the
start point $P$.  

Further evidence for seeing summoning tasks as characterising fundamental
features of relativistic quantum theory was given by Hayden and May \cite{hayden2016summoning},
who considered tasks in which a request is made at precisely 
one from a pre-agreed set of call points $\{c_1 , \ldots , c_n \}$;
a request at $c_i$ requires the state to be produced at the
corresponding return point  $r_i \succ c_i$.
They showed that, if the start point $P$ is in the causal past of
all the call points, then the task is possible if and only if 
no two causal diamonds $D_i = \{ x : r_i \succeq x \succeq c_i \}$
are spacelike separated.   That is, the task is possible unless 
the no-cloning and no-superluminal-signalling principles directly
imply its impossibility.   Wu et al. have presented a more efficient
code for this task \cite{wu2018efficient}.   
Another natural type of summoning task allows any number of calls
to be made at call points, requiring that the state be produced
at any one of the corresponding return points.   Perhaps
counter-intuitively, this can be shown to be a strictly harder
version of the task \cite{adlam2016quantum}.  It is possible if and only if the causal
diamonds can be ordered in sequence so that the return point of any 
diamond in the sequence is in the causal future of all call points
of earlier diamonds in the sequence.   Again, the necessity of 
this condition follows (with a few extra steps)
from the no-superluminal-signalling
and no-cloning theorems \cite{adlam2016quantum}. 

The constraints on summoning have cryptographic applications,
since they can effectively force Alice to make choices 
before revealing them to Bob.   Perhaps the simplest and
most striking of these is a novel type of unconditionally secure  
relativistic quantum bit commitment protocol, in which 
Alice sends the unknown state at light speed in one of two directions,
depending on her committed bit \cite{kent2011unconditionally}.  The fidelity bounds on approximate
quantum cloning imply \cite{kent2011unconditionally} the sum-binding security condition
\begin{equation}
p_0 + p_1 \leq 1 + {{2} \over {d+1}} \, , 
\end{equation}
where $d = \dim (H)$ is the dimension of the Hilbert space of the
unknown state and $p_b$ is the probability of Alice successfully unveiling
bit value $b$.  

Summoning is also a natural primitive in distributed quantum
computation, in which algorithms may effectively summon
a quantum state produced by a subroutine to some computation
node that depends on other computed or incoming data. 

From a fundamental perspective, the (im)possibility of various summoning tasks
may be seen either as results about relativistic quantum theory or 
as candidate axioms for a reformulation of that theory. 
They also give a way of exploring and characterising the space of theories 
generalising relativistic quantum theory.
From a cryptographic perspective, we would like to understand 
precisely which assumptions are necessary for the security of 
summoning-based protocols. 
These motivations are particularly strong given the relationship
between no-summoning theorems and no-signalling, since we know 
that quantum key distribution and other protocols can be proven 
secure based on no-signalling principles alone. 
In what follows, we characterise that relationship more precisely,
and discuss in particular the sense in which summoning-based bit
commitment protocols are secure against potentially post-quantum
but non-signalling participants.   These are participants who
may have access to technology that relies on some unknown theory 
beyond quantum theory.   They may thus be able to carry out 
operations that quantum theory suggests is impossible.
However, their technology must not allow them to violate
a no-signalling principle.   Exactly what this implies depends
on which no-signalling principle is invoked.
We turn 
next to discussing the relevant possibilities.  

\section{No-signalling principles and no-cloning}

\subsection{No-signalling principles}

The relativistic no-superluminal-signalling principle states that no
classical or quantum information can be transmitted at faster than
light speed.  We can frame this operationally by considering 
a general physical system that includes agents 
at locations $P_1 , \ldots , P_n$.   Suppose that
the agent at each $P_i$ may freely choose inputs
labelled by $A_i$ and receive outputs $a_i$, which
may probabilistically depend on their and other inputs. 
Let $I = \{ i_1 , \ldots , i_b \}$ and $J = \{ j_1 , \ldots j_c \}$
be sets of labels of points such that $ P_{i_k} \nsucceq P_{j_l}$ 
for all $k \in \{ 1 , \ldots , b \}$ and 
$l \in \{ 1, \ldots , c \}$. 
Then we have 
\begin{eqnarray}
\lefteqn{P( a_{i_1} \ldots a_{i_b} | A_{i_1} \ldots A_{i_b} ) =} \\
&&  p ( a_{i_1}
\ldots a_{i_b}  |  A_{i_1} \ldots A_{i_b}   A_{j_1} \ldots A_{j_c} )
\, . \nonumber
\end{eqnarray}
In other words, outputs are independent of spacelike or future inputs. 

The quantum no-signalling principle for an $n$-partite system composed
of non-interacting subsystems 
states that measurement outcomes on any subset of subsystems are independent
of measurement choices on the others. 
If we label the measurement choices on subsystem $i$ by $A_i$, 
and the outcomes for this choice by $a_i$, then we have 
\begin{equation}\label{nosignalling}
P( a_{i_1} \ldots a_{i_m} | A_{i_1} \ldots A_{i_m} ) = 
P( a_{i_1} \ldots a_{i_m} | A_{1} \ldots A_{n} )  \, .
\end{equation}
That is, so long as the subsystems are non-interacting, the outputs
for any subset are independent of the inputs for the complementary
subset, regardless of their respective locations in space-time.  

The no-signalling principle for a generalised non-signalling theory 
extends this to any notional device with localised pairs of inputs 
(generalising measurement choices) and outputs (generalising
outcomes).   As in the quantum case, this is supposed to hold
true regardless of whether the sites of the localised input/output ports
are spacelike separated.   Generalized non-signalling  
theories may include, for example, the hypothetical bipartite Popescu-Rohrlich 
boxes \cite{popescu1994quantum}, which maximally violate the CHSH inequality, while still
precluding signalling between agents at each site.  

\subsection{The no-cloning theorem}

The standard derivation of the no-cloning theorem \cite{wootters1982single,dieks1982communication} assumes a
hypothetical quantum cloning device.  A quantum cloning device $D$
should take two input states, a general quantum state $\ket{\psi}$ 
and a reference state $\ket{0}$, independent of $\ket{\psi}$.
Since $D$ follows the laws of quantum
theory, it must act linearly.  Now 
we have 
\begin{equation}
D \ket{\psi} \ket{0} = \ket{\psi} \ket{\psi} \, , \qquad 
D \ket{\psi'} \ket{0} = \ket{\psi'} \ket{\psi'} \, ,
\end{equation}
for a faithful cloning device, for any states $\ket{\psi}$ and 
$\ket{\psi'}$. 
Suppose that $\braket{\psi'}{\psi} = 0$
and that $\ket{\phi} = a \ket{\psi} + b \ket{\psi'} $ is normalised.
We also have  
\begin{equation}
D \ket{\phi} \ket{0} = \ket{\phi} \ket{\phi} \, , 
\end{equation}
which contradicts linearity.

To derive the no-cloning theorem without appealing to linearity, 
we need to consider quantum theory as embedded within a more 
general theory that does not necessarily respect linearity. 
We can then consistently consider a hypothetical post-quantum
cloning device $D$ which accepts quantum states 
$\ket{\psi}$ and $\ket{0}$ as inputs, and produces two
copies of $\ket{\psi}$ as outputs: 
\begin{equation}
D \ket{\psi} \ket{0} = \ket{\psi} \ket{\psi} \, .
\end{equation}
We will suppose that the cloning device functions in
this way independent of the history of the input state.
We will also suppose that it does not violate any
other standard physical principles: in particular, 
if it is applied at $Q$ then it
does not act retrocausally to influence the outcomes
of measurements at earlier points $P \prec Q$. 

We can now extend the cloning device to a bipartite device comprising 
a maximally entangled quantum state, with a standard
quantum measurement device at one end, and the cloning
device followed by a standard quantum measurement device
at the other end.  This extended device accepts classical inputs
(measurement choices) and produces classical outputs
(measurement outcomes) at both ends.   

If we now further assume that the joint output probabilities for 
this extended device, for any set of inputs, are 
independent of the locations of its components, then we can derive a contradiction with
the relativistic no-superluminal signalling principle. 
First suppose that the two ends are timelike separated, 
with the cloning device end at point $Q$ and the other
end at point $P \prec Q$.  
A complete projective measurement at $P$ then 
produces a pure state at $Q$ in any standard
version of quantum theory.   
The cloning device then clones this pure state. 
Different measurement choices at $P$ produce 
different ensembles of pure states at $Q$.
These ensembles correspond to the same mixed state
before cloning, but to distinguishable mixtures
after cloning.   The measurement device at $Q$ can
distinguish these mixtures.    
Now if we take the first end to be at a point $P'$
spacelike separated from $Q$, by hypothesis the 
output probabilities remain unchanged.  
This allows measurement choices at $P'$ to be 
distinguished by measurements at $Q$, and so
gives superluminal signalling \cite{gisin1998quantum}. 

It is important to note that
the assumption of location-independence is not logically necessary,
nor does it follow from the relativistic no-superluminal-signalling
principle alone.   Assuming that quantum states collapse in some well
defined and localized way as a result of measurements, one can consistently 
extend relativistic quantum theory to include hypothetical 
devices that read out a classical description of the local
reduced density matrix at any given point, i.e. the local quantum
state that is obtained by taking into account
(only) collapses within the past light cone
\cite{kent2005nonlinearity}.  
This means that measurement events at $P$, which we take to induce
collapses, are taken into account by the readout device at $Q$ if and only
if $P \prec Q$.  
Given such a readout device, one can certainly clone pure quantum states.
The device behaves differently, when applied to a subsystem of 
an entangled system, depending on whether the second subsystem
is measured inside or outside the past light cone of the 
point at which the device is applies.  It thus does not 
satisfy the assumptions of the previous paragraph.

The discussion above also shows that quantum theory augmented by cloning or readout devices is not a generalized non-signalling
theory.   For consider again a maximally entangled bipartite quantum
system with one subsystem at space-time point $P$ and the
other at a space-like separated point $P'$.   Suppose that
the Hamiltonian is zero, and that the subsystem at $P'$ 
will propagate undisturbed to point $Q \succ P$.  
Suppose that a measurement device may carry out any
complete projective measurement at $P$, and that 
at $Q$ there is a cloning device followed by another
measurement device on the joint (original and cloned)
system.  
As above,  different measurement choices at $P$ produce 
different ensembles of pure states at $Q$, which correspond to the same mixed state
before cloning, but to distinguishable mixtures
after cloning.   The measurement device at $Q$ can
distinguish these mixtures.    
The output (measurement outcome) probabilities at $Q$ thus depend on the 
inputs (measurement choices) at $P$, contradicting 
Eqn. (\ref{nosignalling}). 
Assuming that nature is described by a generalized non-signalling
theory thus gives another reason for excluding cloning or readout
devices, without assuming that their behaviour is location-independent.

In summary, neither the no-cloning theorem nor cryptographic 
security proofs based on it can be derived purely from consistency
with special relativity.   They require further
assumptions about the behaviour of post-quantum devices available to
participants or adversaries.   Although this was noted when
cryptography based on the 
no-signalling principle was first introduced \cite{barrett2005no}, it perhaps deserves re-emphasis.  

On the positive side, given these further assumptions, one can prove
not only the no-cloning theorem, but also quantitative bounds on the
optimal fidelities attainable by approximate cloning devices for
qubits \cite{gisin1998quantum} and qudits \cite{navez2003cloning}.
In particular, one can show \cite{navez2003cloning} that any 
approximate universal cloning device that produces output states
$\rho_0$ and $\rho_1$ given a pure input qudit state $\ket{\psi}$
satisfies the fidelity sum bound 
\begin{equation}\label{acf}
\braopket{\psi}{\rho_0}{\psi}  + 
\braopket{\psi}{\rho_1}{\psi}  
 \leq 1 + {{2} \over {d+1}} \, . 
\end{equation}
It is worth stressing that (with the given assumptions) this bound
applies for any approximate cloning strategy, with any 
entangled states allowed as input.     

\section{Summoning-based bit commitments and no-signalling}

We recall now the essential idea of the flying qudit bit commitment 
protocol presented in Ref. \cite{kent2011unconditionally}, in its idealized form. 
We suppose that space-time is Minkowski and that both parties, the committer (Alice) and the recipient (Bob),
have arbitrarily efficient technology, limited
only by physical principles. 
In particular, we assume they both can carry out error-free quantum 
operations instantaneously and can send classical and quantum
information
at light speed without errors.    
They agree in advance on some
space-time point $P$, to which they have independent secure access,
where the commitment will commence. 

We suppose too that Bob can keep a state secure from Alice 
somewhere in the past of $P$ and arrange to transfer it to her 
at $P$.  Alice's operations on the state can then 
be kept secure from Bob unless and until she chooses to return information
to Bob at some point(s) in the future of $P$.        
We also suppose that Alice can send any relevant states
at light speed in prescribed directions along secure quantum
channels, either by ordinary physical transmission or by
teleportation.   

They also agree on a fixed inertial reference 
frame, and two opposite spatial directions within that frame.    
For simplicity we neglect the $y$ and $z$ coordinates and take the
speed of light $c=1$.   Let $P = (0,0)$ be the origin in the 
coordinates $(x,t)$ and the opposite two spatial directions be defined
by the vectors $v_0 = (-1 , 0 )$ and $v_1 = (1,0 )$.   

Before the commitment begins, Bob generates a random pure qudit
$\ket{\psi} \in {\cal C}^d$.
This is chosen from the uniform distribution, and encoded in some
pre-agreed physical system.   Again idealizing, we assume the
dimensions of this system are negligible, and treat it as pointlike.  
Bob keeps his qudit secure until the point $P$, where he gives it to 
Alice.    To commit to the bit $i \in \{0,1 \}$, Alice sends the state $\ket{\psi}$ along a 
secure channel at light speed in the direction $v_i$.
That is, to commit to $0$, she sends the qudit along the line
$L_0 = \{ (-t, t) , t > 0 \}$; to commit to $1$, she sends it along
the line $L_1 = \{ (t, t) , t > 0 \}$. 

For simplicity, we suppose here that Alice directly transmits the
state along a secure channel.  
This allows Alice the possibility of unveiling her commitment at any point along the transmitted light ray.
To unveil the committed bit $0$, Alice returns $\ket{\psi}$ to Bob at some point $Q_0$ on $L_0$;
to  unveil the committed bit $1$, Alice returns $\ket{\psi}$ to Bob at some point $Q_1$ on $L_1$. 
Bob then tests that the returned 
qudit is $\ket{\psi}$ by carrying out the projective measurement
defined by $P_{\psi} = \ket{\psi} \bra{\psi}$ and its complement $(I -
P_{\psi})$. 
If he gets the outcome corresponding to $P_{\psi}$, he accepts the commitment as honestly
unveiled; if not, he has detected Alice cheating.

Now, given any strategy of Alice's at $P$, there is an optimal state
$\rho_0$ she
can return to Bob at $Q_0$ to maximise the chance of passing his test
there,
i.e. to maximize the fidelity $\braopket{\psi}{\rho_0}{\psi}$. 
There is similarly an optimal state $\rho_1$ that she can return
at $Q_1$, maximizing $\braopket{\psi}{\rho_1}{\psi}$. 
The relativistic no-superluminal-signalling principle implies
that her ability to return $\rho_0$ at $Q_0$ cannot depend on
whether she chooses to return $\rho_1$ at $Q_1$, or vice versa.
Hence she may return both (although this violates the protocol).
The bound (\ref{acf}) on the approximate cloning fidelities implies that
\begin{equation}
\braopket{\psi}{\rho_0}{\psi}  + 
\braopket{\psi}{\rho_1}{\psi}  
 \leq 1 + {{2} \over {d+1} }\, . 
\end{equation}
Since the probability of Alice successfully unveiling the
bit value $b$ by this strategy is 
\begin{equation}
p_b = \braopket{\psi}{\rho_b}{\psi}  \, , 
\end{equation}
this gives the sum-binding security condition for the bit 
commitment protocol
\begin{equation}
p_0 + p_1 \leq 1 + {{2} \over {d+1} }\, .
\end{equation}
Recall that the bound (\ref{acf}) follows from the relativistic 
no-superluminal-signalling condition together with the 
location-independence assumption for a device based on a hypothetical
post-quantum cloning device applied to one subsystem
of a bipartite entangled state.   Alternatively, it follows from
assuming that any post-quantum devices operate within a 
generalized non-signalling theory.  The bit commitment security
thus also follows from either of these assumptions.  

\subsection{Security against post-quantum no-superluminal-signalling adversaries?}

It is a strong assumption that any post-quantum theory
should be a generalized non-signalling theory satisfying 
Eqn. (\ref{nosignalling}). 
So it is natural to ask whether cryptographic security
can be maintained with the weaker assumption that
other participants
or adversaries are able to carry out quantum operations 
and may also be equipped with post-quantum devices, but do
not have the power to signal superluminally. 
It is instructive to understand the limitations of this 
scenario for protocols between mistrustful parties capable of
quantum operations, such as the bit
commitment protocol just discussed. 

The relevant participant here is Alice, who begins with a 
quantum state at $P$ and may send components along the 
lightlike lines $PQ_0$ and $PQ_1$.   Without loss of generality
we assume these are the only components: she could also send
components in other directions, but relativistic
no-superluminal-signalling
means that they cannot then influence her states at $Q_0$
or $Q_1$.  

At any points $X_0$ and $X_1$ on the lightlike lines, before 
Alice has applied any post-quantum devices, the approximate
cloning fidelity bound again implies that 
fidelities of the respective components $\rho_{X_0}$ 
and $\rho_{X_1}$ satisfy
\begin{equation}
\braopket{\psi}{\rho_{X_0}}{\psi}  + 
\braopket{\psi}{\rho_{X_1}}{\psi}  
 \leq 1 + {{2} \over {d+1} }\, . 
\end{equation}

Now, if Alice possesses a classical no-superluminal-signalling device, 
such as a Popescu-Rohrlich box,
with input and output ports at $X_0$ and $X_1$, and her
agents at these sites input
classical information uncorrelated with their quantum states, 
she does not alter the fidelities $\braopket{\psi}{\rho_{X_i}}{\psi}$.
Any subsequent operation may reduce the fidelities, but cannot
increase them. 
More generally, any operation involving the quantum states
and devices with purely classical inputs and outputs cannot increase the 
fidelity sum bound (\ref{acf}).
To see this, note that any such operation
could be paralleled by local operations within quantum theory 
if the two states were held at the same point, since 
hypothetical classical devices with separated pairs of input
and output ports are replicable by ordinary probabilistic classical devices when the ports
are all at the same site.   

We need also to consider the possibility that Alice
has no-superluminal signalling devices with quantum inputs and outputs. 
At first sight these may seem unthreatening. 
For example, while a device that sends the quantum input
from $X_0$ to the output at $X_1$ and vice versa
would certainly make the protocol insecure -- Alice
could freely swap commitments to $0$ and $1$ -- such 
a device would be signalling.  

However, suppose that Alice's agents each have 
local state readout devices, which
give Alice's agent at $X_0$ a  classical description of the density
matrix $\rho_{X_0}$ and 
Alice's agent at $X_1$ a  classical description of the density
matrix $\rho_{X_1}$.   
Suppose also that Alice has carried out an approximate universal cloning at $P$, 
creating mixed states  $\rho_{X_0}$ 
and $\rho_{X_1}$ of the form
\begin{equation}
\rho_{X_i} = p_i \ket{\psi} \bra{\psi} + (1 - p_i ) I \, , 
\end{equation} 
where $0 < p_i < 1 $.
This is possible provided that $p_0 + p_1 \leq 1 + {{2} \over {d+1}
}$.  From these, by applying their readout devices, 
each agent can infer $\ket{\psi}$
locally.   Alice's outputs at $X_i$ have no dependence on the
inputs at $X_{\bar{i}}$.   Nonetheless, this hypothetical 
process would violate the security of the commitment to
the maximum extent possible,   since it would give $p_0 + p_1 =2$. 

To ensure post-quantum security, our post-quantum theory thus need assumptions -- like those spelled out earlier --
that directly preclude state readout devices and other violations of
no-cloning bounds.  

\section{Discussion} 
Classical and quantum relativistic bit commitment protocols have
attracted much interest lately, both because of their theoretical
interest and because advances in theory
\cite{chailloux2017relativistic} and practical implementation
\cite{lunghi2013experimental, liu2014experimental, verbanis201624}
suggest that relativistic cryptography may be in widespread use
in the forseeable future. 

Much work on these topics is framed in models in which
two (or more) provers communicate with one (or more) 
verifiers, with the provers being unable to communicate
with one another during the protocol.   Indeed, one round
classical relativistic bit commitment protocols give
a natural physical setting in which two (or more) separated provers communicate
with adjacent verifiers, with the communications timed so 
that the provers cannot communicate between the commitment and
opening phases.  The verifiers are also typically unable to
communicate, but this is less significant given the form of 
the protocols, and the verifiers are sometimes considered as
a single entity when the protocol is not explicitly relativistic. 

Within the prover-verifier model, it has been shown that
no single-round two-prover classical bit commitment protocol can be 
secure against post-quantum provers who are equipped 
with generalized no-signalling devices \cite{fehr2015multi}.
It is interesting to compare this result with the
signalling-based security proof for the protocol discussed above.

First, of course, the flying qudit protocol involves quantum rather
than classical 
communication between ``provers'' (Alice's agents) and ``verifiers''
(Bob's agents).

Second, as presented, the flying qudit protocol involves three agents for each party.
However, a similar secure bit commitment protocol can be
defined using just two agents apiece.   For example, 
Alice's agent at $P$ could retain the qudit, while 
remaining stationary in the given frame, to commit to $0$,
and send it to Alice's agent at $Q_1$ (as before) to 
commit to $1$.   They may unveil by returning the qudit
at, respectively, $(0,t)$ or $(t,t)$.  In this variant, the commitment is 
not secure at the point where the qudit is received, 
but it becomes secure in the causal future of $(t/2, t/2)$.  

Third, the original flying qudit protocol illustrates a possibility in 
relativistic quantum cryptography that is not motivated 
(and so not normally considered) in standard multi-prover
bit commitment protocols.  This is that, while there are 
three provers, communication between them in some directions is possible (and
required) during the protocol.   
Alice's agent at $P$ must be able to send the quantum state
to either of the agents at $Q_0$ or $Q_1$; indeed, a
general quantum strategy requires her to send quantum
information to both. 

Fourth, the security proof of the flying qudit
protocol can be extended to generalised no-signalling theories.
However, the protocol is not secure if the committer 
may have post-quantum devices that respect the no-superluminal
signalling principle, but are otherwise unrestricted.    
Security proofs require stronger assumptions, such as
that the commmitter is restricted to devices allowed by
a generalized non-signalling theory. 

The same issue arises considering the
post-quantum security of quantum key distribution protocols
\cite{barrett2005no}), which are secure if a post-quantum
eavesdropper is restricted by a generalised no-signalling
theory but not if she is only restricted by the no-superluminal-signalling
principle.  
One distinction is that quantum key distribution is a protocol
between mutually trusting parties, Alice and Bob, whereas bit
commitment protocols involve two mistrustful parties.
It is true that quantum key distribution still involves mistrust, 
in that Alice and Bob mistrust
the eavesdropper, Eve.   However, if one makes the standard cryptographic
assumption that Alice's 
and Bob's laboratories are secure, 
so that information about operations within them cannot propagate to
Eve, one can justify a stronger no-signalling principle
\cite{barrett2005no}.
Of course, the strength of this justification may be questioned,
given that one is postulating unknown physics that could imply
a form of light speed signalling that cannot be blocked. 
But in any case, the justification is not available when one considers protocols
between two mistrustful parties, such as bit commitment, and 
wants to exclude the possibility that one party (in our case
Alice) cannot exploit post-quantum operations within her
own laboratories (which may be connected, forming a single
extended laboratory).   

Our discussion assumed a background Minkowski space-time, but
generalizes to other space-times with standard causal structure,
where the causal relation $\prec$ is a partial ordering.
Neither standard quantum theory nor the usual form of the no-superluminal
signalling principle hold in space-times with closed time-like curves,
where two distinct points $P$ and $Q$ may obey both $P \prec Q$
and $Q \prec P$.   Formulating consistent theories in this context
requires further assumptions (see for example
Ref. \cite{bennett2009can} for one analysis). 
The same is true of superpositions of space-times with indefinite
causal order \cite{oreshkov2012quantum}.    
We leave investigation of these cases for future work. 
 
\vskip 10pt

{\bf Acknowledgments} \qquad This work was partially supported by 
UK Quantum Communications Hub grant no. EP/M013472/1 and by 
Perimeter Institute for Theoretical Physics. Research at Perimeter
Institute is supported by the Government of Canada through Industry
Canada and by the Province of Ontario through the Ministry of Research
and Innovation.   I thank Claude Cr\'epeau and Serge Fehr for 
stimulating discussions and the Bellairs Research Institute for 
hospitality.  
	
\bibliographystyle{unsrtnat}
\bibliography{summoning.signalling}{}

\end{document}